\documentclass[]{krakproc}
\usepackage{graphicx}
\begin{document}

\title{Magnetized Turbulence}
\author{Mordecai-Mark Mac Low}
\institute{Dept. of Astrophysics, American Museum of Natural History,
  79th Street at Central Park W., New York, NY 10024-5192, USA}
\markboth{M.-M. Mac Low}{Magnetized Turbulence}

\maketitle

\begin{abstract}
Several topics in the theory of magnetized turbulence are reviewed
with application to star formation and the interstellar medium.  The
density, pressure, and temperature distribution in a turbulent
interstellar medium is described in comparison to a medium dominated
by thermal instability.  The derivation of the empirical theory for
the energy spectrum of hydrodynamic and MHD turbulence is outlined,
and comparisons are given to numerical models for the magnetized
case. Next discussed is how density fluctuations modify observations
of velocity fluctuations in line centroids, coincidentally
cancelling the effects of projection smoothing. The analytic
description of velocity structure functions as a function of the
dimensionality of the dissipative structures is then covered.
Finally the question of whether turbulence can support against
gravitational collapse is addressed.  Turbulence both prevents and
promotes collapse, but its net effect is to inhibit collapse, so it
does not trigger star formation in any significant sense.

\end{abstract}

\section{Observations}

For more than half a century, evidence for the turbulent nature of the
interstellar medium (ISM) has been known (von Weisz\"acker
\cite{vw43,vw51}; M\"unch \cite{m58}). Armstrong, Rickett, \& Spangler
(\cite{ars95}) compiled observations of the ionized ISM that appear to
show that the magnetized gas has a fluctuation power spectrum
consistent with Kolmogorov (\cite{k41}) turbulence over scales ranging
from 100~pc to small fractions of an AU.  

Other observations pointing to the presence of supersonic, magnetized
turbulence in the ISM include the presence of superthermal linewidths
in the diffuse ionized medium, the warm and cold neutral gas, and
molecular clouds. Line centroid variations also show correlations with
separation suggestive of turbulence.  Magnetic field strengths are
measured using the Zeeman effect or, more recently, the
Chandrasekhar-Fermi (\cite{cf53}) method (Heitsch et al.\ \cite{h01};
Ostriker, Stone, \& Gammie \cite{osg01}; Matthews, Fiege, \&
Moriarty-Schieven \cite{mfm02}; Crutcher et al.\ \cite{c04}). Small
scale fluctuations are measured using refractive and diffractive
pulsar scintillation, and fluctuations in rotation measure and
dispersion measure.

This review briefly discusses the density and temperature structure of
turbulent gas, as a counterpoint to multi-phase models of the ISM.
Analytic descriptions of magnetized turbulence are then treated, as
well as numerical tests of those descriptions.  The observational
consequences of density fluctuations on the line centroid velocity
fluctuation spectrum are briefly noted.  Finally the question of
whether magnetized turbulence can support against gravitational
collapse is reviewed.  Topics that are not covered that fall under the
general topic include magnetized turbulence generated by the
magnetorotational instability, dynamos and field structure in
turbulent flow, and diffusion of cosmic rays or tracers such as heavy
elements.

\section{Density and Temperature Structure}

Classical theories of the interstellar medium have emphasized the
thermal phases produced by the structure of the cooling curve as a
function of temperature (Field, Goldsmith, \& Habing \cite{fgh69}; for a
modern example see Wolfire et al.\ \cite{w95}). At constant pressure,
isolated points of stable equilibrium can be found on the
pressure-density plane, giving phase structure to the ISM.  However, turbulent
flows generate continuous distributions of pressure (Mac Low et al.\ 
\cite{m04}; Avillez \& Breitschwerdt \cite{ab04}), a point also
emphasized by Breitschwerdt and Avillez in these proceedings.
Although thermal instability still acts in such flows, broad regions
on the pressure-density plane end up occupied, rather than isolated
phases. 

\section{Scaling Relations in Magnetized Turbulence}

Our understanding of magnetized turbulence rests on the foundations of
the Kolmogorov (\cite{k41}) theory for the energy spectrum of
incompressible hydrodynamical turbulence. This theory requires two
assumptions: that energy input at large scale and energy dissipation
at small scale balance, and that energy transfer between spatial
scales occurs only between neighboring scales in the intervening
inertial range. Or, to quote Richardson,
\begin{quote}
  Big whirls have little whirls, which feed on their velocity, \\
  and little whirls have lesser whirls, and so on to viscosity.
\end{quote}
The energy of incompressible turbulence $E \propto v^2/2$, and the
crossing time of a perturbation over any scale $\ell$ is $\Delta t =
\ell / v_{\ell}$, where $v_{\ell}$ is the average velocity difference
across structures at that scale.  The assumptions of constant flow
between neighboring scales requires that the energy flow through any
scale
\begin{equation}
  \label{eq:edot}
  \dot{E}_{\ell} \propto \frac{E_{\ell}}{\Delta t} \propto
  \frac{v_{\ell}^3}{\ell}  
\end{equation}
must be constant.  Thus, the velocity must scale as 
\begin{equation}
  \label{eq:vscale}
  v_{\ell} \propto \ell^{1/3}.
\end{equation}
In wavenumber space, $v_k \propto k^{-1/3}$.  We can
derive the energy spectrum in wavenumber space $E(k)$ in either 1D or
3D.  In 1D, the energy in some portion of the inertial range
\begin{equation}
  \label{eq:1d}
  E  = \int E(k) dk \propto v_k^2.
\end{equation}
Substituting the scaling relation derived from equation~\ref{eq:edot},
$k E(k) \propto k^{-2/3}$, so $E(k) \propto k^{-5/3}$.  In 3D, $E =
\int E(k) d^3k$, so $k^3 E(k) \propto k^{-2/3}$, and $E(k) \propto
k^{-11/3}$.  

A theory of incompressible MHD turbulence was proposed by Iroshnikov
(\cite{i63}) and Kraichnan (\cite{k65}) based on the idea that energy
transfer between scales would still be local, but transferred by
interactions between oppositely directed Alfv\'en waves rather than
vortices. Isotropy was still assumed in their derivation, however, as
was a small ratio of fluctuating to mean magnetic field strength.
The interaction time between Alfv\'en waves at any scale $\delta t =
\ell /v_A$ is determined by their propagation speed $v_A^2 = B^2 /
(4\pi\rho)$. The change in any energy from each interaction
\begin{equation}
  \label{eq:ikde1}
  \Delta E \propto \frac{dv^2}{dt} \delta t \propto v_{\ell}
  \dot{v}_{\ell} \delta t.
\end{equation}
The perturbation crossing time $\Delta t$ determines the change in
velocity $\dot{v}_{\ell} = v_{\ell} / \Delta t \propto v_{\ell}^2 /
\ell$, so
\begin{equation}
  \label{eq:ikde2}
  \Delta E \propto \frac{v_{\ell}^3}{\ell} \frac{\ell}{v_A} \propto
  \frac{v_{\ell}^3}{v_A}. 
\end{equation}
The time required for energy to pass through an incoherent cascade can
be shown to be 
\begin{equation}
  \label{eq:casc}
  t_c \propto \left(\frac{v_{\ell}^2}{\Delta E}\right)^2 \delta t
  \propto \frac{\ell v_A}{v_{\ell}^2},
\end{equation}
so the energy flow $\dot{E} \propto v_{\ell}^2 / t_c \propto
v_{\ell}^4 / (\ell v_A)$.  If the energy flow through each scale is
constant, then $v_{\ell} \propto \ell^{1/4}$ in this case.  The energy
spectrum in 1D can then be derived analogously to the hydrodynamic
spectrum, yielding $E(k) \propto k^{-3/2}$.

However, MHD turbulence in the presence of a mean field with energy of
the same order as the flow is not isotropic.  Goldreich \& Sridhar
(\cite{gs95,gs97}) argued that the anisotropy can be taken into
account by treating the energy cascade along parallel lines as
Alfv\'en wave interactions with time scale $t_{\parallel} = \ell/v_A$,
following Iroshnikov and Kraichnan, but treating the perpendicular
cascade with hydrodynamic interactions that merely stir field lines
with time scale $t_{\perp} = \ell/v_{\ell}$.  If the cascades in the
two directions maintain what is referred to as critical balance and
proceed at equal rates, so that $t_{\parallel} = t_{\perp}$, then
$k_{\parallel} v_A = k_{\perp} v_k$. From the hydrodynamic cascade, we
know that $v_k \propto k_{\perp}^{-1/3}$, so the anisotropy of the
turbulence can be derived to be scale dependent: $k_{\parallel}
\propto k_{\perp}^{2/3}$.  The parallel cascade is predicted to still
be an Alfv\'en cascade with $E_{\parallel}(k) \propto
k_{\parallel}^{3/2}$, while the perpendicular cascade looks more like
a Kolmogorov cascade with $E_{\perp}(k) \propto k_{\perp}^{5/3}$.  The
1D spectrum is predicted to be $E(k) \propto k^{5/3}$.

Maron \& Goldreich (\cite{mg01}) confirmed the anisotropy predicted,
but found a 1D spectral index of 3/2.  Cho \& Vishniac (\cite{cv00}),
and Maron, Cowley, \& McWilliams (\cite{mcm04})
on the other hand, do find the predicted 5/3 spectral index. The
difference appears to be the strength of the initial field.  If the
fluctuations in the magnetic field are small compared to the mean
field, the local Alfv\'en cascade dominates, as found by Maron \&
Goldreich (\cite{mg01}).  If the fluctuations are larger, though, the
perpendicular hydrodynamic cascade takes effect, as seen by Cho \&
Vishniac (\cite{cv00}) and Maron et al.\ (\cite{mcm04}).
Finally, if the fields are weak compared to the flow, as in the case
of the turbulent dynamo, the magnetic cascade is no longer local, and
the power spectrum of the field is dominated by high-$k$ modes, while
the velocity power spectrum remains dominated by the driving scale
(e.g.\ Maron et al.\ \cite{mcm04}).

\section{Observations of Velocity Spectra}

One diagnostic of turbulence is the fluctuation in velocity centroid
from point to point across a map.  This is a projection of the 3D
velocity fluctuation spectrum into 2D.  The 3D spectral exponent is
$\Theta =1/3$ in the incompressible hydrodynamical case (eq.\ 
\ref{eq:vscale}), while for shock-dominated Burgers turbulence $\Theta
= 1/2$ (Burgers \cite{b74}).  The projection to 2D should reduce the value
of $\Theta$ by 0.5 in the incompressible case (von Hoerner
\cite{vh51}, Brunt et al.\ \cite{b03}). However, the observed value of
the 2D velocity centroid fluctuation spectrum is close to 0.5
(e.g. Brunt \& Mac Low \cite{bm04}), as if no projection smoothing had
occurred.

Supersonic turbulence drives density fluctuations.  Velocity centroids
depend on both velocity fluctuations and density fluctuations, as the
emissivity at any particular velocity along a line of sight depends on
the density at that point.  These density fluctuations add extra power
at small scales, increasing the slope of the velocity centroid
fluctuation spectrum (Lazarian \& Esquivel \cite{le03}; Brunt \& Mac
Low \cite{bm04}).  By a lucky coincidence, the density fluctuations in
strongly supersonic turbulence cancel the effects of projection
smoothing rather accurately.  This does not hold for transsonic or
subsonic turbulence however, and there is a transition region in the
mildly supersonic regime where the cancellation is incomplete (Brunt
\& Mac Low \cite{bm04}).

\section{Velocity Structure Functions}

Velocity structure functions offer another way of statistically
characterizing turbulence that has proved tractable to analytic
description.  The structure function of order $p$ at scale $L$ is
\begin{equation}
  \label{eq:str-fcn}
  S_p(L) = \langle |v(x + L) - v(x)|^p \rangle.
\end{equation}
The structure function of order two can be directly related to the
energy spectrum $E(k)$.  In the inertial range of isotropic
turbulence, structure functions scale as a power of the length scale
$S_p(L) \propto L^{\zeta(p)}$.  The behavior of the power-law
exponents $\zeta(p)$ as a function of the order $p$ is argued to
depend on the dimension $D$ of dissipative structures in the
turbulence, along with the scaling exponents for $v(L) \propto
L^{\Theta}$ and $t(L)\propto L/v \propto L^{\Delta}$ (She \&
L\'ev\^eque \cite{sl94}; Dubrulle \cite{d94}).  The form of the
scaling is
\begin{equation}
  \label{eq:zeta}
  \zeta(p) = \Theta(1-\Delta)p + C\left(1-\Sigma^{p \Theta}\right),
\end{equation}
where the codimension $C = 3-D$, and $\Sigma = 1 - \Delta/C$.  In
Kolmogorov turbulence, equation~(\ref{eq:vscale}) shows that $\Theta =
1/3$, so $\Delta = 2/3$.

In incompressible hydrodynamical turbulence, the dissipative
structures are linear vortex tubes with dimension $D=1$ (She \&
L\'ev\^eque \cite{sl94}).  M\"uller \& Biskamp (\cite{mb00}) showed
that incompressible MHD behaves as if its dissipative structures have
dimension $D=2$; they suggested that these structures are current
sheets. Compressible MHD has also been examined.  Boldyrev
(\cite{b02}) and Boldyrev, Nordlund, \& Padoan (\cite{bnp02}) suggest
that the dissipative structures have $D=2$ and are primarily
shocks. Padoan et al. (\cite{p04}) show that in super-Alfv\'enic
but subsonic to transsonic turbulence $1 < D < 2$ depending on the
sonic Mach number ${\cal M}_{\rm rms}$. However, Vestuto et al.\
(\cite{v03}) raise a note of caution, as they find steeper power laws
for energy spectra than would be predicted by the velocity structure
function of order 2 under this formalism.

\section{Gravitational Support}

An important question for understanding star formation is whether
magnetized turbulence can provide support against gravitational
collapse.  In order to provide significant support, it must act like a
pressure in all directions, and it must last longer than a free-fall
time.  Neither of these requirements is clearly met by supersonic
turbulence in molecular clouds.

Furthermore, hydrostatically supported objects do not form easily in
supersonic turbulent flows.  This is because an isothermal object
in hydrostatic equilibrium requires a balancing external pressure to confine
it. Too high a pressure sends it into collapse, while too low a
pressure allows it to again expand. V\'azquez-Semadeni, et al.\ (\cite{vs04})
have demonstrated numerically that these two paths are taken by
virtually all density fluctuations in a magnetized turbulent flow,
leaving very few objects that might be subject to ambipolar diffusion
and loss of magnetic support over long time periods.  Rather, the star
formation rate appears to be limited by the necessity of assembling
supercritical regions by moving gas along field lines in order to
increase the mass-to-flux ratio, and by the global limitation of
collapse by the turbulence, when that is effective.

If turbulence decays in less than a free-fall time, it will not
provide significant support against gravitational collapse.  For
several decades, Arons \& Max (\cite{am75}) was interpreted to mean that
magnetic fields could substantially reduce the decay rate of
turbulence.  In the late 1990's, three groups demonstrated numerically
that even magnetized turbulence decays quickly (Mac Low et al.\ 
\cite{m98}, Stone, Ostriker, \& Gammie \cite{s98}, Padoan \& Nordlund
\cite{pn99}). Mac Low et al.\ (\cite{m98}) showed that supersonic,
trans-Alfv\'enic or super-Alfv\'enic turbulence decays as $E(t)
\propto t^{-1}$, with a resolution study ranging from 32$^3$ to
256$^3$ zones. 

There are a couple of exceptions to this quick decay known, although
their astrophysical relevance remains to be demonstrated.
Biskamp \& M\"uller (\cite{bm99}) used 512$^3$
simulations of incompressible MHD to show that $E \propto t^{-1}$ for
flows with magnetic helicity $H = \vec{A} \cdot \vec{B} = 0$, but that
$E \propto t^{-1/2}$ for flows with significant helicity.  Unbalanced
cascades of Alfv\'en waves have been shown to decay more slowly by
Maron \& Goldreich (\cite{mg01}) and Cho, Lazarian, \& Vishniac
(\cite{clv02}).  However, to reduce decay significantly, Alfv\'en wave
fluxes must be as much as an order of magnitude higher in one
direction than the opposite.

Mac Low (\cite{m99,m03}) estimates that the dissipation rate for
isothermal, supersonic turbulence is
\begin{equation} \label{eqn:dissip}
\dot{e} \simeq -(1/2)\rho v_{\rm rms}^3/L_d, 
\end{equation}
where $L_d$ is the driving scale.
The dissipation time for turbulent kinetic energy
\begin{equation} \label{eqn:disstime}
\tau_d = e / \dot{e} \simeq L/v_{\rm rms}, 
\end{equation}
which is just the crossing time for the turbulent flow across the
driving scale (Elmegreen \cite{e00}).

Stone et al.\ (\cite{s98}) and Mac Low (\cite{m99}) showed that
supersonic turbulence decays in less than a free-fall time under
molecular cloud conditions, regardless of whether it is magnetized or
unmagnetized.  The hydrodynamical result agrees with the
high-resolution, transsonic, decaying models of Porter \& Woodward
(\cite{pw92}) and Porter, Pouquet, \& Woodward (\cite{pw94}).  Mac Low
(\cite{m99}) showed that the formal dissipation time $\tau_{\rm d} =
e/\dot{e}$ scaled in units of the free fall time $t_{\rm ff}$ is
\begin{equation} \label{eqn:decay}
\tau_{\rm d}/\tau_{\rm ff} = \frac{1}{4 \pi \xi} \left(\frac{32}{3}\right)^{1/2}
\frac{\kappa}{{\cal M}_{\rm rms}} \simeq \,3.9 \,\frac{\kappa}{{\cal
M}_{\rm rms}},
\end{equation}
where $\xi = 0.21/\pi$ is the Kolmogorov energy-dissipation
coefficient derived by Mac Low (\cite{m99}), ${\cal M}_{\rm rms} =
v_{\rm rms}/c_{\rm s}$ is the rms Mach number of the turbulence, and
$\kappa$ is the ratio of the driving wavelength to the Jeans
wavelength $\lambda_{\rm J}$.  In molecular clouds, ${\cal M}_{\rm
  rms}$ is typically observed to be of order 10 or higher.  If the
ratio $\kappa < 1$, as is probably required to maintain gravitational
support (L\'eorat, Passot, \& Pouquet \cite{l90}), then even strongly
magnetized turbulence will decay long before the cloud collapses, and
not markedly retard the collapse.

Supersonic turbulence both prevents and promotes collapse. It can
provide support against collapse on scales larger than the driving
scale. Classical theories treat small-scale turbulence as an isotropic
pressure with effective sound speed $c_{s, {\rm eff}} = (c_s + \langle
v^2 \rangle / 3)^{1/2}$ (Chandrasekhar \cite{c51}; von Weisz\"acker
\cite{vw51}).  By this argument, supersonic turbulence {\em increases}
the Jeans mass to $M_{J, {\rm eff}} = (\pi/G)^{3/2} \rho^{-1/2} c_{s,
  {\rm eff}}^3$.  However, supersonic flows drive strong density
fluctuations on scales below the driving scale.  Positive density
fluctuations in isothermal turbulence reach densities $\rho' = \rho
{\cal M}_{\rm rms}^2$.  In these density enhancements, the Jeans mass
{\em decreases} to $M_{J, {\rm eff}} = (\pi/G)^{3/2} \rho'^{-1/2}
c_{s, {\rm eff}}^3$. If we express everything in terms of velocity,
however, 
\begin{equation}
  \label{eq:jeans-v}
  M_{J, {\rm eff}} = (\pi/G)^{3/2} \rho'^{-1/2} c_{s, {\rm eff}}^3
  \propto  \frac{c_s}{v} \left( c_s + \langle
v^2 \rangle / 3 \right)^{3/2} \propto v^2.
\end{equation}
Even though turbulence can locally promote collapse, on average it
{\em inhibits} collapse when compared to the same conditions absent
turbulence.  Thus, although turbulence may give the appearance of
locally triggering collapse and star formation, it is globally
preventing star formation, acting as a counterbalance to gravitational
instability.  Turbulence controls star formation in opposition to
gravity, not in alliance with gravity.



\vskip 0.1in
I thank J. Maron \& A. Schekochihin for useful discussions, the
conference organizers for partial support of my attendance, and
J. Pulte for hospitality during the writing of the proceedings. Further
support was provided by NSF grant AST99-85392 and NASA grant NAG5-10103.
Extensive use was made of the NASA Astrophysics Data System in
preparation of this work.


\begin{thebibliography}{}
\bibitem[1995]{ars95}
Armstrong, J. W., Rickett B. J., \& Spangler, S. R. 1995, ApJ, 443,
209
\bibitem[1975]{am75}
Arons, J., \& Max, C. E. 1975, ApJ (Letters), 196, L77
\bibitem[2004]{ab04}
Avillez, M. A., \& Breitschwerdt, D. 2004, A\&A, 425, 899
\bibitem[1999]{bm99} Biskamp, D., \& M\"uller, W.-C. 1999,
  Phys. Rev. Lett. 83, 2195
\bibitem[2002]{b02} Boldyrev, S. 2002, ApJ, 569, 841
\bibitem[2002]{bnp02} Boldyrev, S., Nordlund, \AA., \& Padoan,
  P. 2002, Phys. Rev. Lett., 89, 1102
\bibitem[2003]{b03} Brunt, C. M., Heyer, M. H., V\'azquez-Semadeni, E.,
  \& Pichardo, B. 2003, ApJ, 595, 824
\bibitem[2004]{bm04} Brunt, C. M., \& Mac Low, M.-M. 2004, ApJ, 604, 196
\bibitem[1974]{b74} Burgers, J. M. 1974, The Nonlinear Diffusion
  Equation (Dordrecht: Reidel) 
\bibitem[1951]{c51} Chandrasekhar, S. 1951, Proc. R. Soc. London A, 210, 26
\bibitem[1953]{cf53}
Chandrasekhar, S., \& Fermi, E. 1953, ApJ, 118, 113
\bibitem[2002]{clv02} Cho, J., Lazarian, A., \& Vishniac, E. T. 2002,
  ApJ, 564, 291
\bibitem[2000]{cv00} Cho, J., \& Vishniac, E. T. 2000, ApJ, 539, 273
\bibitem[2004]{c04}
Crutcher, R. M., Nutter, D. J., Ward-Thompson, D., \& Kirk,
J. M. 2004, ApJ, 600, 279
\bibitem[1994]{d94} Dubrulle, B. 1994, Phys. Rev. Lett., 73, 959
\bibitem[2000]{e00}
Elmegreen, B.\ G., 2000, ApJ, 530, 277  
\bibitem[1969]{fgh69}
Field, G. B., Goldsmith, D. W., \& Habing, H. J. ApJ (Letters), 155, L149
\bibitem[1995]{gs95} Goldreich, P.\ \& Sridhar, S. 1995, ApJ, 438, 763
\bibitem[1997]{gs97} Goldreich, P.\ \& Sridhar, S. 1997, ApJ, 485, 680
\bibitem[2001]{h01}
Heitsch, F., Zweibel, E. G., Mac Low, M.-M., Li, P., \& Norman,
M. L. 2001, ApJ, 561, 800
\bibitem[1963]{i63}
Iroshnikov, P. S. 1963, Astron. Zh., 40, 742 (Engl.\ transl.: 1964,
Sov. Astron., 7, 566) 
\bibitem[1941]{k41} Kolmogorov, A. N. 1941, Dokl. Akad. Nauk SSSR, 32,
  141 (Engl.\ transl.: 1958, Amer. Math. Soc. Transl., Ser. 2, 8, 87.)
\bibitem[1965]{k65}
Kraichnan, R. 1965, Phys. Fluids, 8, 1385
\bibitem[2003]{le03} Lazarian, A., \& Esquivel, A. 2003, ApJ, 592, L37
\bibitem[1990]{l90} L\'eorat, J., Passot, T. \& Pouquet, A. 1990,
  MNRAS, 243, 293.
\bibitem[1999]{m99} Mac Low, M.-M. 1999, ApJ, 524, 169
\bibitem[2003]{m03} Mac Low, M.-M. 2003, in Simulations of
magnetohydrodynamic turbulence in astrophysics, eds. T. Passot \&
E. Falgarone, Lecture Notes in Physics, 614, 182
\bibitem[2004]{m04} Mac Low, M.-M., Balsara, D. S., Kim, J., \&
  Avillez, M. A. 2004, ApJ, submitted (astro-ph/0410734)
\bibitem[1998]{m98} Mac Low, M.-M., Klessen, R. S., Burkert, A., \&
Smith, M. D. 1998, Phys. Rev. Lett., 80, 2754 
\bibitem[2004]{mcm04} Maron, J., Cowley, S., \& McWilliams, J. 2004,
  ApJ, 603, 569
\bibitem[2001]{mg01} Maron, J., \& Goldreich, P. 2001, ApJ, 554, 1175
\bibitem[2002]{mfm02} Matthews, B. C., Fiege, J. D., \&
  Moriarty-Schieven, G. 2002, ApJ, 569, 304
\bibitem[2000]{mb00} M\"uller, W.-C., \& Biskamp, D.  2000, Phys. Rev.
  Lett., 84, 475
\bibitem[1958]{m58} M\"unch, G. 1958, Rev. Mod. Phys., 30, 1035
\bibitem[2001]{osg01}
Ostriker, E. C., Stone, J. M., \& Gammie, C. F. 2001, ApJ, 546, 980
\bibitem[2004]{p04} Padoan, P., Jiminez, R., Nordlund, \AA., \&
  Boldyrev, S. 2004, Phys. Rev. Lett., 92, 191102
\bibitem[1999]{pn99}
Padoan, P., \& Nordlund, \AA. 1999, ApJ, 526, 279 
\bibitem[1992]{pw92} Porter, D.\ H., \& Woodward, P. R. 1992,
  Phys. Rev. Lett., 68, 3156
\bibitem[1994]{pw94} Porter, D.\ H., Pouquet, A. \& Woodward, P. R.
  1994, Phys. Fl., 6, 2133
\bibitem[1994]{sl94} She, Z., \& L\'ev\^eque, E. 1994, Phys. Rev.
  Lett., 72, 336
\bibitem[1998]{s98} 
Stone, J.\ M., Ostriker, E. C., \& Gammie, C. F. 1998, ApJ, 508, L99
\bibitem[2004]{vs04} V\'azquez-Semadeni, E., Kim, J., Shadmehri, M., \&
  Ballesteros-Paredes, J. 2004, ApJ, in press (astro-ph/0403134)
\bibitem[2001]{v03}
Vestuto, J. G., Ostriker, E. C., \& Stone, J. M. 2003, ApJ, 590, 858
\bibitem[1951]{vh51} von Hoerner, S. 1951, Z. Astrophys., 30, 17
\bibitem[1943]{vw43}
von Weizs\"acker, C. F. 1943, Z. Astrophys., 22, 319
\bibitem[1951]{vw51}
von Weizs\"acker, C. F. 1951, ApJ, 114, 165
\bibitem[1995]{w95}
Wolfire, M., Hollenbach, D., McKee, C. F., Tielens, A. G. G. M., \&
Bakes, E. L. O. 1995, ApJ, 443, 152
\end{thebibliography}
\end{document}